\documentstyle[12pt]{article}
\newlength{\aivwidth}   
\newlength{\tmpwidth}   

\newcommand{\phrd}[1]{Phys.\ Rev.\ {\bf D#1}}

\newcommand{\nphb}[1]{Nucl.\ Phys.\ {\bf B#1}}

\newcommand{\phlb}[1]{Phys.\ Lett.\ {\bf B#1}}

\newcommand{\zphc}[1]{Z.\ Phys.\ {\bf C#1}}

\newcommand{\ijmpha}[1]{Int.\ J.\ Mod.\ Phys.\ {\bf A#1}}

\newcommand{\be}{\begin{equation}}
\newcommand{\ee}{\end{equation}}
\newcommand{\bea}{\begin{eqnarray}}
\newcommand{\eea}{\end{eqnarray}}
\newcommand{\ba}{\begin{array}}
\newcommand{\ea}{\end{array}}
\newcommand{\eref}[1]{(\ref{#1})}
\newcommand{\nn}{\nonumber\\}

\newcommand{\suu}{$\rm SU(2)_L\times U(1)_Y$}
\newcommand{\lag}{{\cal L}}
\newcommand{\ep}{\epsilon}
\newcommand{\edd}{\ep_{\scriptscriptstyle DD}}
\newcommand{\pa}{\partial}

\newcommand{\cw}[1]{\cos^{#1}\theta_W}
\newcommand{\ldd}{\lag_{DD}}
\newcommand{\lddt}{\lag_{DD,2}}
\newcommand{\ww}{$e^+e^-\to W^+W^-$}
\newcommand{\tr}{\mbox{tr}}

\begin{document}

\title{Nonstandard Quartic Self Interactions\\
of Electroweak Vector Bosons\\
within a Gauge Invariant\\Higher-Derivative Term}
\author{Carsten Grosse-Knetter\thanks{E-Mail: knetter@physw.uni-bielefeld.de}
\\[5mm]Universit\"at Bielefeld\\
Fakult\"at f\"ur Physik\\33501 Bielefeld\\Germany}
\date{BI-TP 94/01\\hep-ph/9401201\\January 1994}
\maketitle

\begin{abstract}
A locally \suu-invariant term of dimension six with effective interactions
of the electroweak gauge and Higgs fields is examined.
This term also contains higher derivatives of the fields.
It yields nonstandard quartic gauge-boson
self-interactions but neither nonstandard
cubic self-interactions nor tree-level
effects on presently measurable quantities.
\end{abstract}


Effective (nonstandard)
interactions among the electroweak vector bosons $W^\pm$, $Z$ and
$\gamma$ have been investigated very intensively with respect to experimental
tests of the vector-boson self-interactions in future collider experiments
(LEP~II, NLC, LHC).

In recent works the nonstandard self couplings are parametrized through
\suu\ gauge invariant Lagrangians with linearly realized symmetry
\cite{llr,buwy,ruj,haze1,haze2,gkks}, i.e. effective Lagrangians of
the type
\be \lag_{eff}=\lag_{SM}+\sum_{n>4}\sum_i \frac{\ep_i^{(n)}}{\Lambda^{n-4}}
\lag^{(n)}_i,\label{leff}\ee
where $\lag_{SM}$ is the (dimension-four) standard-model Lagrangian
and the $\lag_i^{(n)}$ are locally \suu -invariant effective interaction terms
of (higher) dimension $n$.
$\Lambda$ is the scale of the assumed new physics beyond
the effective theory and the $\ep_i^{(n)}$ are coupling constants.
Usually one considers only dimension-six terms because terms of higher
dimension are suppressed by higher negative powers of $\Lambda$.

The complete list of dimension-six effective interaction terms is given
in \cite{llr}. These terms can be classified in three cathegories:
\begin{itemize}
\item Terms which contain fermionic couplings or two-gauge-boson interactions.
These affect four-fermion amplitudes at the tree level and thus would imply
strong deviations of LEP~I results from the standard-model predictions.
Therefore the empirical limits on the coupling constants of these terms are
very strict \cite{ruj,haze2,gkks,grwi}.
\item Terms that contain nonstandard gauge boson self-interactions but
no fermionic couplings or two-gauge-boson couplings. These terms
affect LEP~I observables only indirectly
through loop effects. However, as a consequence
of the linearly realized gauge invariance, the loop contributions
of such (nonrenormalizable) effective interactions to observable
quantities are very small; they depend only logarithmically on the
cut-off $\Lambda$ \cite{ruj,haze1,haze2}. This means that the indirect bounds
on the corresponding coupling constants that can be deduced from LEP~I results
are not very strict \cite{ruj,haze1,haze2}. The vector-boson self-interactions
contained in these terms thus have to be tested directly in future experiments
in which vector bosons are produced.
\item Terms which contain neither fermionic couplings
nor gauge-boson self-interactions. They only yield
nonstandard interactions involving the Higgs boson. These terms are
not ruled out from LEP~I data, too, however they are also not very interesting
with respect to tests of the vector-boson self-interactions.
\end{itemize}
There is one dimension-six term that belongs to the second cathegory and
which has not been considered in the literature so far, namely the term
\be \lag_{DD}= -\frac{1}{2}\mbox{tr}\left[(D_\mu D^\mu \Phi)^\dagger
(D_\nu D^\nu \Phi)\right]. \label{ldd}\ee
In \eref{ldd} the following notation is used:
\bea
D_\mu \Phi&=&\partial_\mu\Phi+igW_\mu\Phi-\frac{i}{2}g'\Phi\tau_3B_\mu,\nn
\Phi&=&\frac{1}{\sqrt{2}}((v+H){\bf 1}+i\varphi_i\tau_i),\nn
W_\mu&=&\frac{1}{2}W_{\mu i}\tau_i,\label{not}\eea
with $v$ being the vacuum expectation value of the scalar fields, $H$ the Higgs
field, $\varphi_i$ the pseudo-Goldstone fields and $W_{\mu i}$ and $B_\mu$
the gauge fields.
In this paper I will discuss the term $\lag_{DD}$; i.e. I will investigate
the effective Lagrangian
\be \lag_{eff}=\lag_{SM}+\frac{\edd}{\Lambda^2}\ldd \label{leffdd}.\ee

First one notes that the term \eref{ldd} contains higher derivatives
of the scalar fields. It is well known that higher-order Lagrangians, in
general, imply unphysical effects (unphysical degrees of freedom, unbound
energy from below, no analytic limit for $\ep\to 0$) \cite{hide}.
However, these unphysical effects are absent if an {\em effective\/}
higher-order Lagrangian is considerd  \cite{high},
i.e. a Lagrangian which is assumed to be the
low-energy approximation of ``new physics''. In this case, a
higher-order Lagrangian can formally be treated like
a first-order one, especially
the Feynman rules can be obtained in the usual way from the effective
Lagrangian \cite{high,hle}. Other dimension-six terms
with higher derivatives that, however, effect LEP~I observables at the
tree level have been examined in \cite{grwi}.

Like any effective interaction term with higher derivatives, \eref{ldd}
can be rewritten as a (physically equivalent)
term without higher derivatives by applying the
standard-model equations of motion \cite{high}. (Actually, such an
application of the equations of motion corresponds to a field transformation
\cite{eom} which becomes a canonical transformation within the
Hamiltonian formalism for higher-order Lagrangians \cite{high}.)
However, for practical calculations such a reduction is not very useful,
because then a term like \eref{ldd} becomes a complicated expression
from which the physical interpretation
of this term is not very obvious. (See \cite{ruj,gkks} for examples.)
It is more convenient to examine the term \eref{ldd} in its
simple form with higher derivatives given above.

The above statement that $\lag_{DD}$ has no tree-level effects on
four-fermion amplitudes is not strictly correct because the effective
Lagrangian \eref{leffdd} contains a term
\be -\frac{\edd}{2}
\frac{M_W^2}{\Lambda^2}\left[2(\pa_\mu W^{+\mu})(\pa_\nu W^{-\nu})+
\frac{1}{\cw{2}}(\pa_\mu Z^\mu)(\pa_\nu Z^\nu)\right] \label{quad} \ee
which affects the propagators of the massive gauge bosons. These become
in the unitary gauge ($\varphi_i=0$)
\be -i\frac{\displaystyle g^{\mu\nu}-\left(1-\frac{\Lambda^2}
{\edd M_B^2}\right)
\frac{p^\mu p^\nu}{p^2-\frac{\Lambda^2}{\edd}}}{p^2-M_B^2}, \qquad
M_B=M_W,M_Z.\label{wprop}\ee
However, if one considers Feynman graphs with a virtual
gauge-boson that is coupled to light fermions
with mass $m_f$, the deviations of the corresponding amplitudes
from the standard model induced by the propagator \eref{wprop}
are proportional to
\[\frac{m_f}{M_B}\ll 1\]
and thus negligible. Therefore the term $\lag_{DD}$ has de facto no tree-level
effect on LEP~I observables and thus it
cannot be ruled out due to present empirical
data. It should be noted that the propagator \eref{wprop} contains
in addition to the pole at $p^2=M_B^2$ another pole at $p^2=\frac{\Lambda^2}
{\edd}$ which could be interpreted as the effect of heavy particles connected
with the assumed new physics beyond the effective Lagrangian \eref{leffdd}.
Similarly, the Higgs propagator deriving from \eref{leffdd}
\be i\frac{1}{\displaystyle p^2-M_H^2-\frac{\edd}{\Lambda^2}p^4}\label{hprop}
\ee
has poles for $p^2=M_H^2+O(\Lambda^{-2})$ and for $p^2=\frac{\Lambda^2}{\edd}
+O(\Lambda^0)$.

One can easily see that \eref{leffdd} contains no nonstandard cubic
gauge-boson self-interactions but the quartic interaction term
\be -\frac{\edd}{2}
g^2 \frac{M_W^2}{\Lambda^2}\left[W_\mu^+W^{-\mu}W_\nu^+W^{-\nu}
+\frac{1}{\cw{2}}W_\mu^+W^{-\mu}Z_\nu Z^\nu +\frac{1}{4\cw{4}}
Z_\mu Z^\mu Z_\nu Z^\nu\right]\label{quart}.\ee
This means that (within gauge invariant dimension-six extensions of
the standard model) the absence of nonstandard cubic interactions
among the vector bosons does not imply standard quartic couplings, i.e.
also the quartic couplings have to be tested in experiments in order
to verify the standard model. Nonstandard quartic couplings can be implied
by the assumed new physics beyond the effective theory through the exchange of
a heavy particle.

Anomalous quartic vector-boson self-interactions that are independent of
the cubic self interactions have been examined (within gauge
noninvariant models) in \cite{kmss,bebo}. If one requires
custodial SU(2) symmetry (i.e. global SU(2) symmetry in the absence of
the $B_\mu$ field), electromagnetic gauge invariance and the absence of
derivatives one can construct two effective quartic interaction terms
\cite{kmss,bebo}, namely \eref{quart} and
\bea -\frac{\ep_{\scriptscriptstyle DD,2}}{2}
g^2 \frac{M_W^2}{\Lambda^2}&&\!\!\!\!\!\!
\bigg[\frac{1}{2}(W_\mu^+W^-_\nu W^{+\mu}W^{-\nu}+
W_\mu^+W^{-\mu}W_\nu^+W^{-\nu})
\nn&&\!\!\!\!\!\!+
\frac{1}{\cw{2}}W_\mu^+W^-_\nu Z^\mu Z^\nu+\frac{1}{4\cw{4}}
Z_\mu Z^\mu Z_\nu Z^\nu\bigg].\label{quart2}\eea
However, considering gauge invariant
extensions of the standard model,
there is an important difference between these expressions:
\eref{quart} can be emdedded in a gauge invariant dimension-six
term that does not affect LEP~I observables at the tree level but
\eref{quart2} cannot.
Actually, \eref{quart2} is contained in the term
\be \lag_{DD,2}= -\frac{1}{4}\mbox{tr}\left[(D_\mu D_\nu \Phi)^\dagger
(D^\mu D^\nu \Phi)\right]+\frac{1}{2}\ldd ,
\label{ldd2}\ee
which, however, can be expressed through $\ldd$ \eref{ldd} (after dropping
total derivative terms) according to
\be \lag_{DD,2}=\ldd-\frac{1}{4}(g\lag_{W\Phi}+g^\prime\lag_{B\Phi}-
g^2\lag_{WW}-g^{\prime 2}\lag_{BB}+2gg^{\prime}\lag_{WB}).\label{rel}\ee
The additional terms in \eref{rel} are
\bea
\lag_{W\Phi}&=&i\tr[(D_\mu\Phi)^\dagger W^{\mu\nu}(D_\nu\Phi)]\nn
\lag_{B\Phi}&=&-\frac{1}{2}i\tr[(D_\mu\Phi)^\dagger (D_\nu\Phi)\tau_3]
  B^{\mu\nu}\nn
\lag_{WW}&=&-\frac{1}{2}\tr[\Phi^\dagger W_{\mu\nu}W^{\mu\nu}\Phi]\nn
\lag_{BB}&=&-\frac{1}{8}\tr[\Phi^\dagger \Phi] B_{\mu\nu}B^{\mu\nu}\nn
\lag_{WB}&=&-\frac{1}{4}\tr[\Phi^\dagger W_{\mu\nu}\Phi\tau_3]B^{\mu\nu}
\label{terms}\eea
(with the notation \eref{not}). These terms have been examined elsewhere
\cite{ruj,haze1,haze2,gkks}.
$\lag_{W\Phi}$, $\lag_{B\Phi}$, $\lag_{WW}$ and $\lag_{BB}$ have no
tree-level effects on four-fermion amplitudes
($\lag_{W\Phi}$ and $\lag_{B\Phi}$ contain nonstandard cubic and quartic
vector-boson self-interactions,
$\lag_{WW}$ and $\lag_{BB}$ yield after a redefinition of  the
gauge fields and of the coupling constants only nonstandard couplings of
the Higgs field), but
$\lag_{WB}$ contains a $W_3 B$-mixing term which affects
LEP~I observables at the tree level. Thus, in distinction to $\ldd$ \eref{ldd},
$\lddt$ \eref{ldd2} contradicts LEP~I results. (However,
both expressions, \eref{quart} and
\eref{quart2}, can be embedded in dimension-eight terms that have no tree-level
effects on LEP~I observables \cite{gkks}.)

Since $\ldd$ does not contain cubic vector-boson self-interactions
it even has no tree-level effects on the process \ww\ to
be measured at LEP~II or NLC. However it will yield contributions to
three-gauge-boson production in $e^+e^-$ collisions at NLC
\cite{bebo,www,gksch} and to vector-boson scattering, which can
be realized by scattering vector bosons
emitted by quarks in $pp$ collisions
at LHC \cite{wws,wwns}. These
processes can be used as experimental tests of the
quartic vector-boson self-interactions. The corresponding amplitudes
also get
direct contributions from Higgs-exchange diagrams. Therefore one has to
consider the nonstandard couplings of the Higgs field to the gauge fields
contained in \eref{leffdd}, as well.
Here I list  the additional vector-vector-Higgs
couplings in the unitary gauge ($\varphi_i=0$):
\bea-\frac{\edd}{2} g \frac{M_W}{\Lambda^2}\!\!\!\!\!\!&&\Bigg\{
\left[2(\pa_\mu W^{+\mu})(\pa_\nu W^{-\nu})+
\frac{1}{\cw{2}}(\pa_\mu Z^\mu)(\pa_\nu Z^\nu)\right]H\nn\!\!\!\!\!\!&&{}-
\left[2W_\mu^+ W^{-\mu}+\frac{1}{\cw{2}}Z_\mu Z^\mu\right]
\Box H\nn\!\!\!\!\!\!&&+
2\left[(\pa_\mu W^{+\mu})W^{-\nu}+(\pa_\mu W^{-\mu})W^{+\nu}
+\frac{1}{\cw{2}}(\pa_\mu Z^\mu) Z^\nu\right]\pa_\nu H\Bigg\}.
\eea

Finally it should be noted that an effective gauge theory with the same
vector-boson self-interactions as \eref{leffdd} but without a physical Higgs
boson can be constructed by formally taking the limit $M_H\to\infty$,
i.e. by replacing \cite{apbe,gkko}
\be \Phi=\frac{1}{\sqrt{2}}((v+H){\bf 1}+i\varphi_i\tau_i)\to
\frac{v}{\sqrt{2}}\exp\left(\frac{i\varphi_i\tau_i}{v}\right)\ee
in \eref{leffdd}.
The resulting (chiral)
Lagrangian, however, contains nonpolynonial interactions of
the unphysical scalar fields.

In summary: The term $\ldd$ \eref{ldd} should necessarily
be taken into account
when investigating gauge invariant dimension-six extensions of the electroweak
standard model. The fact that this term contains higher
derivatives does not imply any unphysical effects.
$\ldd$ is the only dimension-six
term which contains
nonstandard quartic vector-boson self-interactions
but no cubic ones  without having tree-level effects on presently measurable
quantities.
These nonstandard quartic couplings have to be tested in future experiments.

{\bf Acknowledgement:} I thank I.~Kuss and D.~Schildknecht for helpful
discussions.




\begin{thebibliography}{99}
\bibitem{llr}C. N. Leung, S. T. Love and S. Rao, \zphc{31} (1986) 433
\bibitem{buwy} W. Buchm\"uller and D. Wyler, \nphb{268} (1986) 621
\bibitem{ruj}A. de R\'{u}jula, M. B. Gavela, P. Hern\'{a}ndez and E. Mass\'{o},
\nphb{384} (1992) 3
\bibitem{haze1}K. Hagiwara, S. Ishihara, R. Szalapski and D. Zeppenfeld,
\phlb{283} (1992) 353;\\
P. Hern\'{a}ndez and F. J. Vegas, Phys. Lett. {\bf B307} (1993) 116
\bibitem{haze2}K. Hagiwara, S. Ishihara, R. Szalapski and D. Zeppenfeld,
\phrd{48} (1993) 2182
\bibitem{gkks}C. Grosse-Knetter, I. Kuss and D. Schildknecht, \zphc{60}
(1993) 375
\bibitem{grwi}B. Grinstein and M. B. Wise, Phys.\ Lett.\ {\bf B265}
(1991) 326
\bibitem{hide}A. Pais and G. Uhlenbeck,
Phys.\ Rev.\ {\bf 79} (1950) 145;\\
C. Bernard and A. Duncan, Phys.\ Rev.\ {\bf D11} (1975) 848;\\
S. W. Hawking, in ``Quantum Field Theory and Quantum
Statistics'', ed.: I. A. Batalin, C. J. Isham and C. A. Vilkovisky
(Hilger, 1987) p.~129;\\
D. A. Eliezer and R. P. Woodard,
Nucl.\ Phys.\ {\bf B325} (1989) 389;\\
J. Z. Simon, Phys.\ Rev.\ {\bf D41} (1990) 3720
\bibitem{high}C. Grosse-Knetter, Bielefeld Preprint
BI-TP 93/29 (1993), hep-ph/9306321
\bibitem{hle}C. Grosse-Knetter,
Bielefeld Preprint BI-TP 93/40 (1993), hep-ph/9308201, to be published
in \phrd{}
\bibitem{eom}D. Barua and S. N. Gupta,
Phys.\ Rev.\ {\bf D16} (1977) 413;\\
H. D. Politzer, Nucl.\ Phys.\ {\bf B172} (1980) 349;\\
G. Sch\"afer, Phys.\ Lett.\ {\bf A100} (1984) 128;\\
T. Damour and G. Sch\"afer, J.\ Math.\ Phys.\ {\bf 32} (1991) 127;\\
H. Georgi, Nucl.\ Phys.\ {\bf B361} (1991) 339;\\
H. Leutwyler, Bern Preprint BUTP-91/26 (1991);\\
C. Arzt, Michigan Preprint UM-TH-92-28 (1992),
hep-ph/9304230;\\
F. Feruglio, \ijmpha{8} (1993) 4937
\bibitem{kmss}
M. Kuroda, J. Maalampi, D. Schildknecht and K. H. Schwarzer,
Nucl. Phys. {\bf B284} (1987) 271
\bibitem{bebo}G. B\'{e}langer and F. Boudjema, \phlb{288} (1992) 201
\bibitem{www}V. Barger, T. Han and R. J. N. Phillips, \phrd{39} (1989) 146;\\
A. Tofighi-Niaki and J. F. Gunion, \phrd{39} (1989) 720
\bibitem{gksch}C. Grosse-Knetter and D. Schildknecht, \phlb{302} (1993)
 309
\bibitem{wws}M. J. Duncan, G. L. Kane and W. W. Repko, \nphb{272} (1986) 517
\bibitem{wwns}J. Bagger, S. Dawson, G. Valencia, \nphb{399} (1993) 364;\\
G. Gounaris and F. M. Renard, \zphc{59} (1993) 143
\bibitem{apbe}T. Appelquist and C. Bernard, \phrd{22} (1980) 200;\\
A. C. Longhitano, \nphb{188} (1981) 118
\bibitem{gkko}C. Grosse-Knetter and R. K\"ogerler, \phrd{48} (1993) 2865
\end{thebibliography}
\end{document}